\documentclass[12pt,a4paper]{article}
\usepackage{epsfig}
\newcommand{\dd}{\mbox{{\rm d}}}

\newcommand{\Lumint}{{\cal L}_{\rm int}}

\begin{document}

\begin{center}
{\Large
{\bf Search for new physics effects in Bhabha scattering 
at $e^+e^-$ linear colliders}}
\\ [0.5cm]
\large
E.S. Kokoulina, L.M. Kurbatova and A.A. Pankov \\[0.5cm]
\normalsize
Gomel State Technical University,\\
Belarus \\[1cm]
\end{center}

\begin{abstract}
\noindent
We study electron-electron contact-interaction searches in the
process  $e^+e^-\to e^+e^-$ at a future 
linear collider with longitudinally polarized
beams, and evaluate the model-independent
constraints on the contact interaction parameters, 
emphasizing the role of beam polarization. 
 
\vspace*{3.0mm}
\end{abstract}

\section{Introduction}

The possibility of constructing high energy polarized electron and
positron beams is considered with great interest with regard to 
the physics programme at a linear collider (LC). Indeed, one 
of the most important advantages of initial beam polarization 
is that one can measure spin-dependent
observables, which represent the most direct probes of the fermion 
helicity dependence of the electroweak interactions. 
Consequently, one would expect a substantial gain in the 
sensitivity to the features of possible
non-standard interactions and, in particular, stringent 
constraints on the individual new coupling constants 
could be derived from the data analysis by 
looking for deviations of the observables from the 
Standard Model (SM) predictions. 

Contact interaction Lagrangians (CI) provide an effective framework 
to account for the phenomenological effects of new dynamics 
characterized by extremely high intrinsic mass scales $\Lambda$, at
the `low' energies $\sqrt s\ll\Lambda$ attainable at current particle 
accelerators. 
The explicit parameterization of the four-fermion quark and lepton 
contact interactions is, {\it a priori}, somewhat arbitrary. In general,  
it must respect $SU(3)\times SU(2)\times U(1)$ symmetry, because   
the new dynamics are active well-beyond the electroweak scale. 
Furthermore, usually one limits to the lowest dimensional operators, 
$D=6$ being the minimum, and neglects higher dimensional operators
that are suppressed by higher powers of $1/\Lambda^2$ and therefore are 
expected to give negligible effects. 

In this note, we consider the effects of the flavor-diagonal, 
helicity conserving ${eeff}$ contact-interaction 
effective Lagrangian \cite{Eichten}
\begin{equation}
{\cal L}_{\rm CI}
=\frac{1}{1+\delta_{ef}}\sum_{\alpha\beta}g^2_{\rm eff}\hskip
2pt\epsilon_{\alpha\beta}
\left(\bar e_{\alpha}\gamma_\mu e_{\alpha}\right)
\left(\bar f_{\beta}\gamma^\mu f_{\beta}\right),
\label{lagra}
\end{equation}
in the Bhabha scattering process 
\begin{equation}
e^++e^-\to e^++e^-, \label{proc}
\end{equation}
at an $e^+e^-$ linear collider  with c.m.\ energy 
$\sqrt s=0.5\hskip 2pt{\rm TeV}$ and polarized electron and positron 
beams. In Eq.~(\ref{lagra}): $\alpha,\beta={\rm L,R}$
denote 
left- or right-handed fermion helicities, $f$ indicates the fermion 
species, so that $\delta_{ef}=1$ for the process (\ref{proc}) under
consideration, and the CI coupling constants are parameterized in
terms of 
corresponding mass scales as 
$\epsilon_{\alpha\beta}={\eta_{\alpha\beta}}/{{\Lambda^2_{\alpha\beta}}}$.
Actually, one assumes $g^2_{\rm eff}=4\pi$ to account for the fact
that 
the interaction would become strong at $\sqrt s\simeq\Lambda$, 
and by convention 
$\vert\eta_{\alpha\beta}\vert=\pm 1$ or $\eta_{\alpha\beta}=0$,
leaving the
energy scales $\Lambda_{\alpha\beta}$ as free, {\it a priori} 
independent parameters.     

For the case of the Bhabha process (\ref{proc}), the effective 
lagrangian interaction in Eq.~(\ref{lagra}) envisages the existence
of three individual, and independent, CI models, contributing to 
individual helicity amplitudes or combinations of them,  
with {\it a priori} free, and non-vanishing, coefficients 
(basically, $\epsilon_{\rm LL},\epsilon_{\rm RR}\ {\rm and}
\ \epsilon_{\rm LR}$ combined with the $\pm$ signs). Correspondingly, 
in principle the most general, and model-independent, analysis 
of the data 
must account for the situation where all four-fermion effective 
couplings defined in Eq.~(\ref{lagra}) are simultaneously allowed in 
the expression for the cross section. 
Potentially, the different CI couplings may interfere and 
substantially weaken the bounds. Indeed, although the different
helicity amplitudes by themselves do not interfere, 
{\it the deviations from the SM} could be positive for one 
helicity amplitude and negative for another, 
so that accidental cancellation might occur in the sought for 
deviations from the SM predictions for the relevant observables. 
It should be highly desirable to apply a general 
(and model-independent) approach to the analysis of 
experimental data, that allows to simultaneously include all 
terms of Eq.~(\ref{lagra}) as independent, non vanishing free 
parameters, and yet to derive separate constraints (or exclusion 
regions) on the values of the CI coupling constants, free from 
potential weakening due to accidental cancellations. 
Such an analysis is feasible with initial beam longitudinal 
polarization and  allows to extract the individual helicity 
cross sections from suitable combinations of measurable 
polarized cross sections and, consequently, to disentangle the 
constraints on the corresponding CI constants.

\section{Sensitivity to CI}

With $P^-$ and $P^+$ the longitudinal polarization  
of the electron and positron beams, respectively, and $\theta$ the angle 
between the incoming and the outgoing electrons in the c.m.\ frame, 
the differential cross section of process (\ref{proc}) at lowest order,
including $\gamma$ and $Z$ exchanges both in the $s$ and $t$ 
channels and the contact interaction (\ref{lagra}), can be written 
in the following form \cite{Schrempp,Renard,Pankov1}: 
\begin{equation}
\frac{\dd\sigma(P^-,P^+)}{\dd\cos\theta}
=(1-P^-P^+)\,\frac{\dd\sigma_1}{\dd\cos\theta}+
(1+P^-P^+)\,\frac{\dd\sigma_2}{\dd\cos\theta}+
(P^+-P^-)\,\frac{\dd\sigma_P}{\dd\cos\theta}.
\label{cross}
\end{equation}
In Eq.~(\ref{cross}): 
\begin{eqnarray}
\frac{\dd\sigma_1}{\dd\cos\theta}&=&
\frac{\pi\alpha^2}{4s}\left[A_+(1+\cos\theta)^2+A_-(1-\cos\theta)^2\right],
\nonumber \\
\frac{\dd\sigma_2}{\dd\cos\theta}&=&
\frac{\pi\alpha^2}{4s}\, 4A_0, 
\nonumber \\
\frac{\dd\sigma_P}{\dd\cos\theta}&=&
\frac{\pi\alpha^2}{4s}\, A_+^P(1+\cos\theta)^2,
\label{sigP}
\end{eqnarray}
with 
\begin{eqnarray}
A_0(s,t)&=&
\left(\frac{s}{t}\right)^2\big\vert 1+g_{\rm R}\hskip 2pt
g_{\rm L}\chi_Z(t)+\frac{t}{\alpha}\,\epsilon_{\rm LR}\big\vert^2,
\nonumber \\
A_+(s,t)&=&
\frac{1}{2}\big\vert 1+\frac{s}{t}+g_{\rm L}^2\hskip 2pt
\left(\chi_Z(s)+\frac{s}{t}\,\chi_Z(t)\right)+
2\frac{s}{\alpha}\,\epsilon_{\rm LL}\big\vert^2 \nonumber \\
&+&
\frac{1}{2}\big\vert 1+\frac{s}{t}+g_{\rm R}^2\hskip 2pt
\left(\chi_Z(s)+\frac{s}{t}\,\chi_Z(t)\right)+
2\frac{s}{\alpha}\,\epsilon_{\rm RR}\big\vert^2, \nonumber \\
A_-(s)&=&
\big\vert 1+g_{\rm R}\hskip 2pt
g_{\rm L}\,\chi_Z(s)+\frac{s}{\alpha}\,\epsilon_{\rm LR}\big\vert^2, 
\nonumber \\
A_+^P(s,t)&=&
\frac{1}{2}\big\vert 1+\frac{s}{t}+g_{\rm L}^2\hskip 2pt
\left(\chi_Z(s)+\frac{s}{t}\,\chi_Z(t)\right)+
2\frac{s}{\alpha}\,\epsilon_{\rm LL}\big\vert^2 \nonumber \\
&-&
\frac{1}{2}\big\vert 1+\frac{s}{t}+g_{\rm R}^2\hskip 2pt
\left(\chi_Z(s)+\frac{s}{t}\,\chi_Z(t)\right)+
2\frac{s}{\alpha}\,\epsilon_{\rm RR}\big\vert^2.
\label{A}
\end{eqnarray}
Here: $\alpha$ is the fine structure constant; $t=-s(1-\cos\theta)/2$ and 
$\chi_Z(s)=s/(s-M^2_Z+iM_Z\Gamma_Z)$,  
$\chi_Z(t)=t/(t-M^2_Z)$ represent the $Z$ propagators in $s$ and $t$
channel,
respectively, with $M_Z$ and $\Gamma_Z$ the mass and width of the $Z$;  
$g_{\rm R}=\tan\theta_W$, $g_{\rm L}=-\cot{2\,\theta_W}$
are the SM right- and left-handed electron couplings of the $Z$, 
with $s_W^2=1-c_W^2\equiv \sin^2\theta_W$.
\par
With both beams polarized, the polarization of each beam can be changed 
on a pulse by pulse basis. This would allow the separate measurement 
of the polarized cross sections for each of the four polarization 
configurations RR, LL, RL and LR, corresponding to the four sets of beam
polarizations 
$(P^-,P^+)=(P_1,P_2)$, $(-P_1,-P_2)$, $(P_1,-P_2)$ and $(-P_1,P_2)$, 
respectively, with $P_{1,2}>0$. Specifically, with the simplifying 
notation $\dd\sigma\equiv\dd\sigma/\dd\cos\theta$:
\begin{eqnarray}
{\dd\sigma_{\rm RR}}&\equiv&{\dd\sigma(P_1,P_2)}=
(1-P_1P_2)\,{\dd\sigma_1}+
(1+P_1P_2)\,{\dd\sigma_2}+(P_2-P_1)\,{\dd\sigma_P},
\nonumber \\
{\dd\sigma_{\rm LL}}&\equiv&{\dd\sigma(-P_1,-P_2)}=
(1-P_1P_2)\,{\dd\sigma_1}+
(1+P_1P_2)\,{\dd\sigma_2}-(P_2-P_1)\,{\dd\sigma_P},
\nonumber \\
{\dd\sigma_{\rm RL}}&\equiv&{\dd\sigma(P_1,-P_2)}=
(1+P_1P_2)\,{\dd\sigma_1}+
(1-P_1P_2)\,{\dd\sigma_2}-(P_2+P_1)\,{\dd\sigma_P},
\nonumber \\
{\dd\sigma_{\rm LR}}&\equiv&{\dd\sigma(-P_1,P_2)}=
(1+P_1P_2)\,{\dd\sigma_1}+
(1-P_1P_2)\,{\dd\sigma_2}+(P_2+P_1)\,{\dd\sigma_P}.
\label{cross4}
\end{eqnarray}
 
To make contact to the experiment we take $P_1=0.8$ and $P_2=0.6$, and 
impose a cut in the forward and backward directions. Specifically, 
we consider the cut angular range $\vert\cos\theta\vert<0.9$ and divide 
it into nine equal-size bins of width $\Delta z=0.2$ ($z\equiv\cos\theta$).
We also introduce the experimental efficiency, $\epsilon$, for
detecting the final $e^+e^-$ pair, and according to the LEP2 experience  
$\epsilon=0.9$ is assumed. 
 
The most natural choice of observables are 
the four, directly measurable, differential event rates integrated 
over each bin:
\begin{equation}
N_{\rm LL},\quad  N_{\rm RR},\quad N_{\rm LR},\quad N_{\rm RL},
\label{obsn}
\end{equation}
where ($\alpha,\beta={\rm L,R}$), and 
\begin{equation}
N_{\alpha\beta}^{\rm bin}=\frac{1}{4}\Lumint\,\epsilon
\int_{\rm bin}(\dd\sigma_{\alpha\beta}/\dd z)\dd z.
\label{n}
\end{equation}
In Eq.~(\ref{n}), $\Lumint$ is the time-integrated luminosity, which is 
assumed to be equally divided among over the four combinations of 
electron and positron beams polarization defined in Eqs.~(\ref{cross4}). 
\par 

The current bounds on $\Lambda_{\alpha\beta}$, 
of the order of several TeV, are such that for the LC c.m.\ energy 
$\sqrt{s}=0.5$~TeV the characteristic suppression factor $s/\Lambda^2$ 
in Eq.~(\ref{A}) is rather strong. Accordingly, we can safely assume 
a linear dependence of the cross sections on the parameters 
$\epsilon_{\alpha\beta}$. In this regard, indirect manifestations of 
the CI interaction (\ref{lagra}) can be looked for, {\it via} deviations 
of the measured observables from the SM predictions, caused by the 
new interaction. 
The reach on the CI couplings, and the corresponding constraints on their 
allowed values in the case of no effect observed, can be estimated 
by comparing the expression of the mentioned deviations with the expected 
experimental (statistical and systematic) uncertainties. 
\par      
To this purpose, assuming the data to be well described by the SM 
($\epsilon_{\alpha\beta}=0$) predictions, i.e., that no deviation is 
observed within the foreseen experimental uncertainty, and in the linear 
approximation in $\epsilon_{\alpha\beta}$ of the observables 
(\ref{obsn}), we apply the method based on the covariance 
matrix:
\begin{eqnarray}
\label{covarv}
V_{kl}&=&\langle({\cal O}_k-\bar{\cal O}_k)
                ({\cal O}_l-\bar{\cal O}_l)\rangle\nonumber \\
&=&
\sum_{i=1}^{4}\left(\delta N_i\right)^2  
\left(\frac{\partial{\cal O}_k}{\partial N_i}\right)
\left(\frac{\partial{\cal O}_l}{\partial N_i}\right)
+\left(\delta \Lumint \right)^2  
\left(\frac{\partial{\cal O}_k}{\partial\Lumint}\right)
\left(\frac{\partial{\cal O}_l}{\partial\Lumint}\right)
\nonumber \\
&+&
\left(\delta P^-\right)^2  
\left(\frac{\partial{\cal O}_k}{\partial P^-}\right)
\left(\frac{\partial{\cal O}_l}{\partial P^-}\right)+
\left(\delta P^+\right)^2  
\left(\frac{\partial{\cal O}_k}{\partial P^+}\right)
\left(\frac{\partial{\cal O}_l}{\partial P^+}\right).
\end{eqnarray}
Here, the $N_i$ are given by Eq.~(\ref{obsn}), so that the statistical 
error appearing on the right-hand-side is given by
\begin{equation}
\delta N_i=\sqrt{N_i}, \label{deltani} \end{equation}
and the ${\cal O}_l=(N_{\rm LL},\quad  N_{\rm RR},\quad N_{\rm LR},
\quad N_{\rm RL}$ are the observables. The second, third and fourth 
terms of the right-hand-side of Eq.~(\ref{covarv}) represent the systematic 
errors on the integrated luminosity $\Lumint$, polarizations $ P^-$ and 
$P^+$, respectively.

\begin{figure}[b]
\refstepcounter{figure}
\label{Fig1}
\addtocounter{figure}{-1}
\begin{center}
\setlength{\unitlength}{1cm}
\begin{picture}(12,7.5)
\put(-3.,0.0)
{\mbox{\epsfysize=8.5cm\epsffile{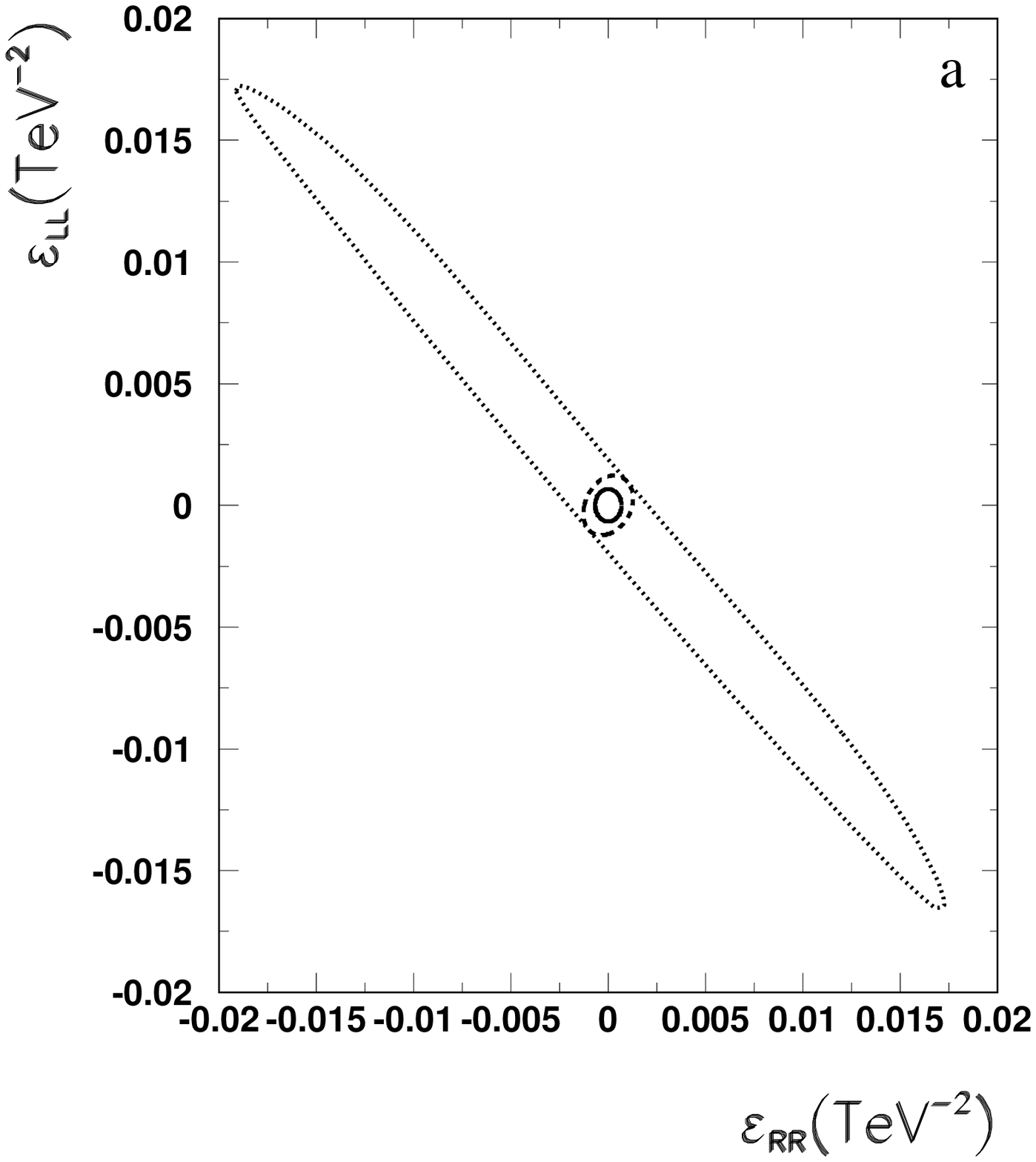}}
 \mbox{\epsfysize=8.5cm\epsffile{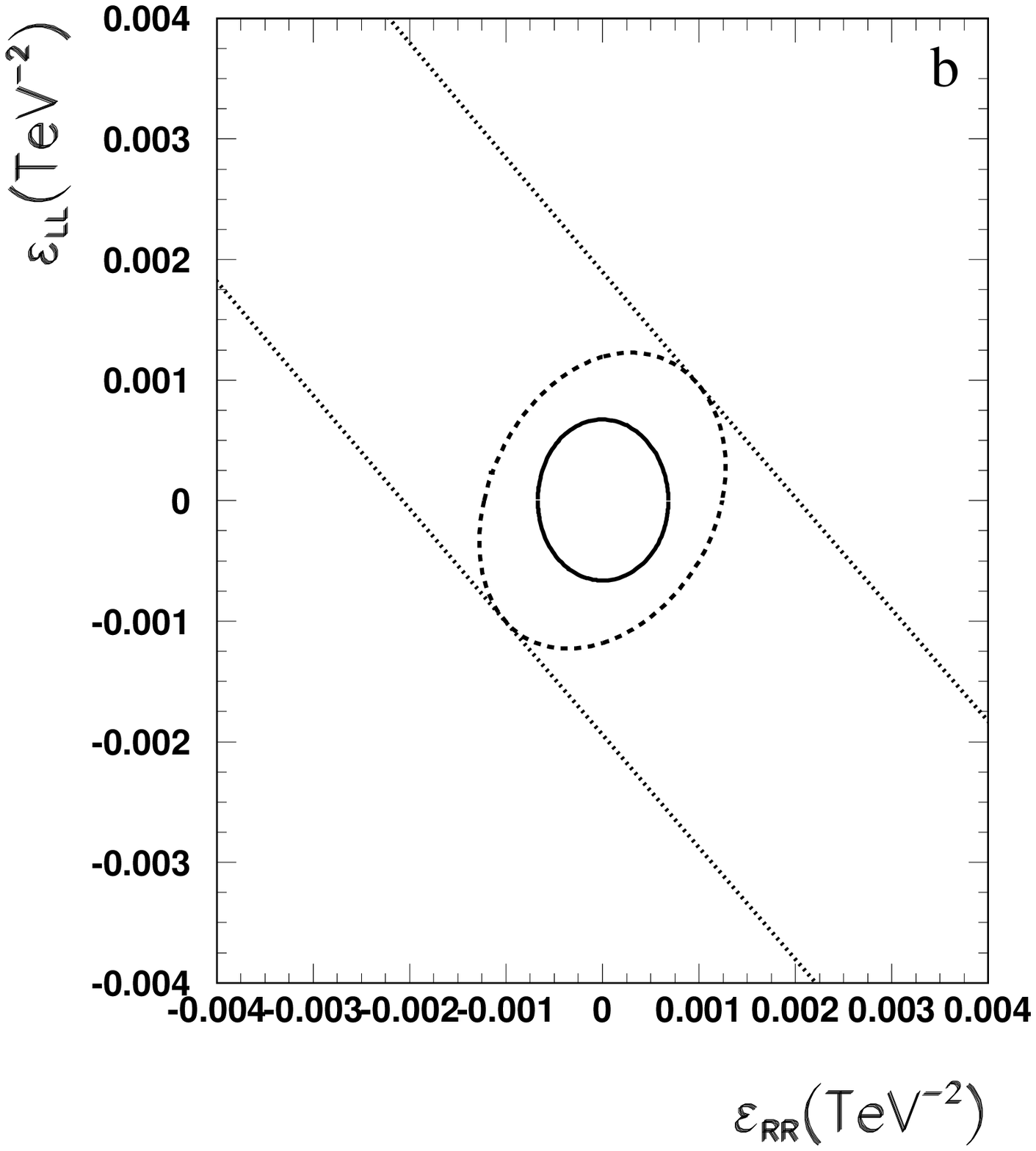}}}
\end{picture}
\vspace*{-5mm}
\caption{
(a) Allowed areas at 95\% C.L. on electron contact 
interaction parameters 
in the plane ($\epsilon_{LL},\epsilon_{LR}$),
obtained from differential polarized cross sections at 
$\sqrt{s}=500$ GeV, $\Lumint=50\ \mbox{fb}^{-1}$,
$\vert P^-\vert=0.8$ and $\vert P^+\vert=0.6$ (solid line),
$\vert P^-\vert=0.8$ and $P^+=0$ (dashed line), and
$P^-=P^+=0$ (dotted line). (b) Magnification 
of allowed domains presented in  Fig.~1a.}
\end{center}
\end{figure}
\begin{figure}[t]
\refstepcounter{figure}
\label{Fig2}
\addtocounter{figure}{-1}
\begin{center}
\setlength{\unitlength}{1cm}
\begin{picture}(12,7.5)
\put(-3.,0.0)
{\mbox{\epsfysize=8.5cm\epsffile{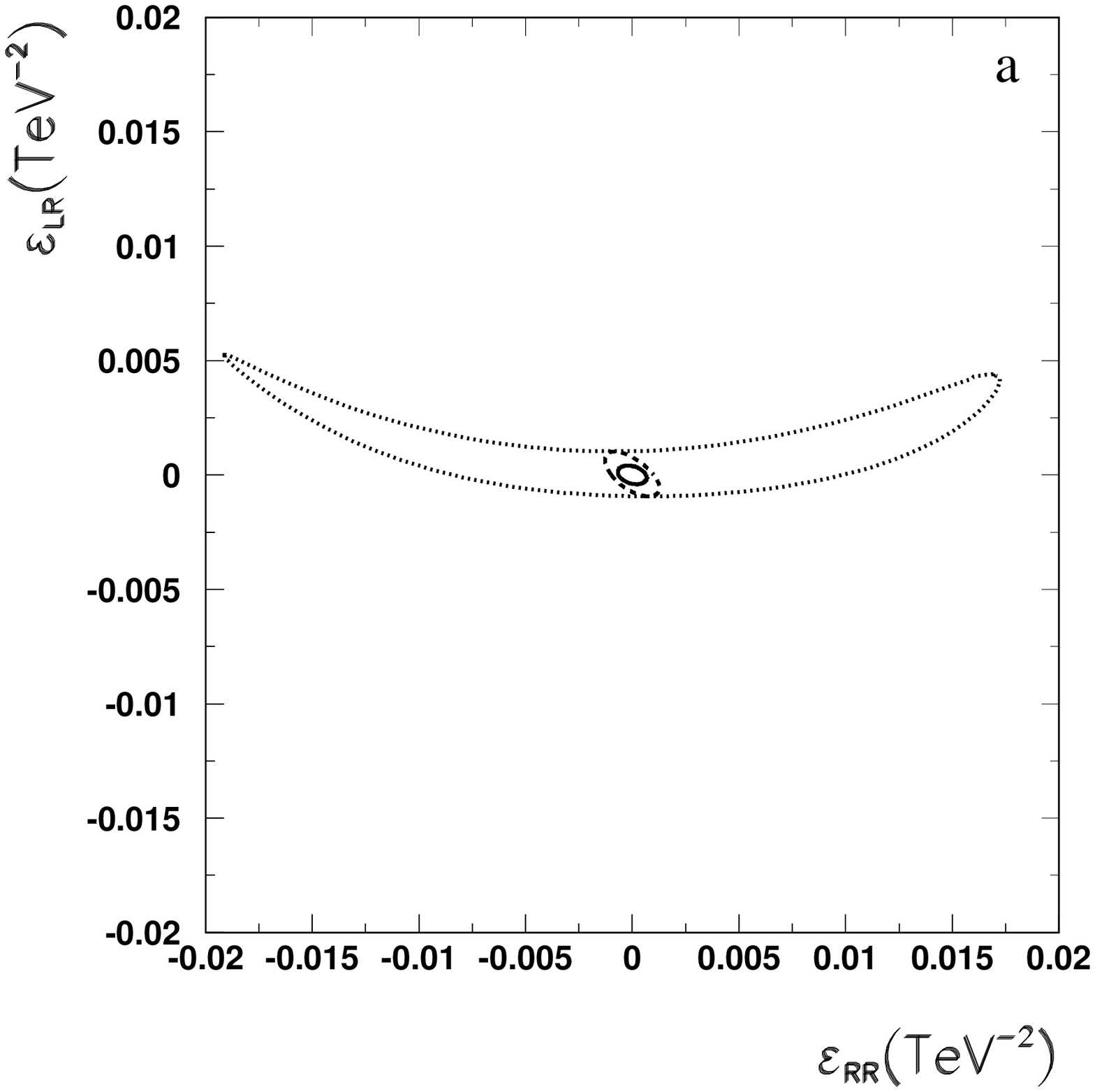}}
 \mbox{\epsfysize=8.5cm\epsffile{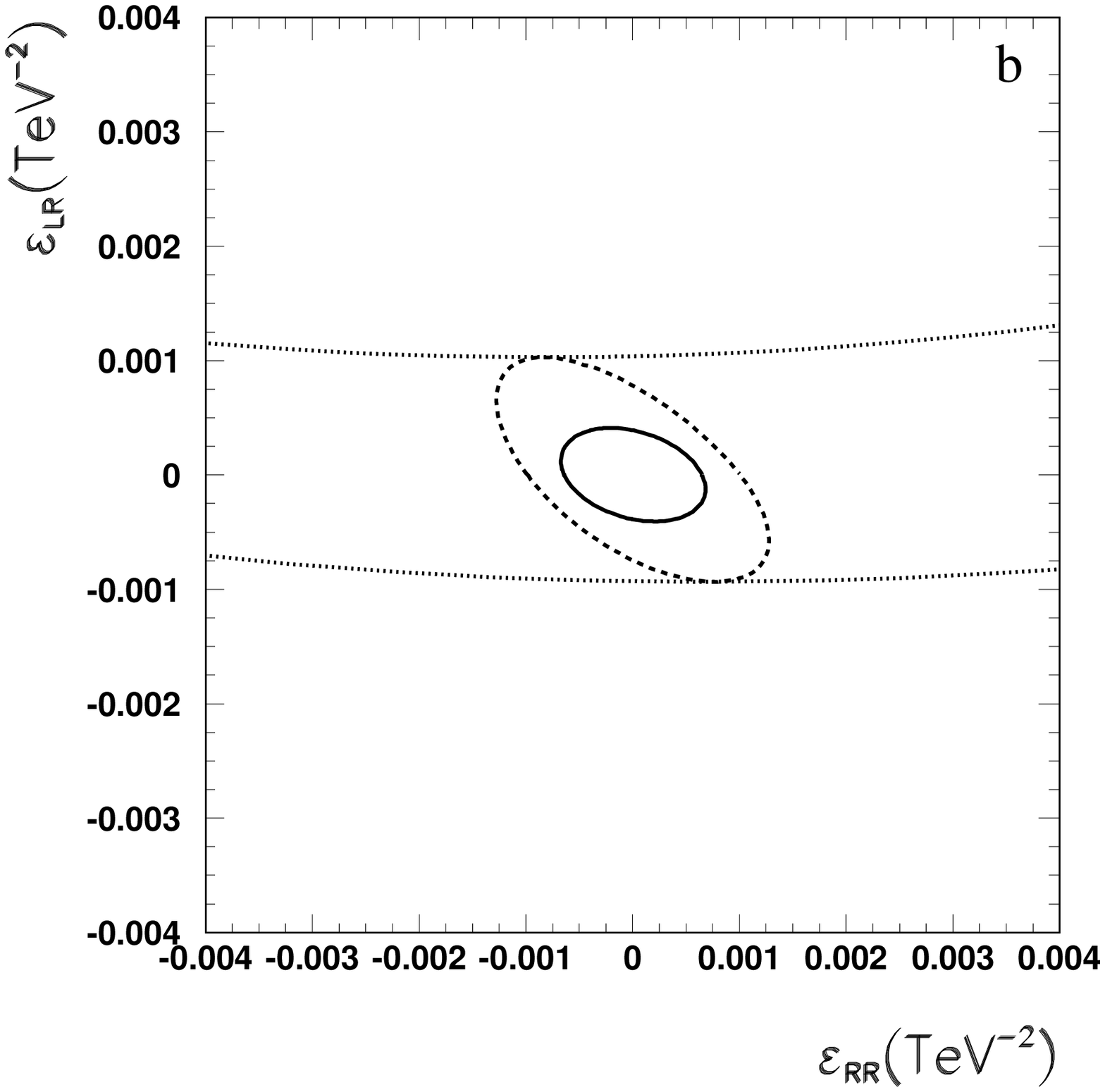}}}
\end{picture}
\vspace*{-5mm}
\caption{Same as in Fig.~1, for ($\epsilon_{LR},\epsilon_{RR}$) 
plane.}
\end{center}
\end{figure}

Defining the inverse covariance matrix $W^{-1}$ as 
\begin{equation}
\left(W^{-1}\right)_{ij}=\sum_{k,l=1}^4\sum_{bins} \left(V^{-1}\right)_{kl}
\left(\frac{\partial{\cal O}_k}{\partial\epsilon_i}\right)
\left(\frac{\partial{\cal O}_l}{\partial\epsilon_j}\right), 
\end{equation}
with 
$\epsilon_i=(\epsilon_{LL},\ \epsilon_{LR},\ \epsilon_{RR})$, 
model-independent allowed domains in the four-dimensional CI parameter space
to 95\% confidence level are obtained from the error contours determined by 
the quadratic form in $\epsilon_{\alpha\beta}$ \cite{Cuypers,Pankov2}:
\begin{equation}
\label{chi2}
\left(\epsilon_{LL}\ \epsilon_{LR}\ \epsilon_{RR}\right) 
W^{-1}
\left( \begin{array}{c}
\epsilon_{LL}\\
\epsilon_{LR}\\
\epsilon_{RR}
\end{array}\right)=7.82.
\end{equation}
The value 7.82 on the right-hand side of Eq.~(\ref{chi2}) corresponds to 
a fit with three free parameters. 
 
The quadratic form (\ref{chi2}) defines a three-dimensional ellipsoid in the 
$\left(\epsilon_{LL},\ \epsilon_{LR},\ \epsilon_{RR}\right)$
parameter space. 
The matrix $W$ has the property that the square roots of the individual   
diagonal matrix elements, $\sqrt{W_{ii}}$, determine the projection 
of the ellipsoid onto the corresponding $i$-parameter axis in the 
three-dimensional space, and has the meaning of the bound at 95\% C.L. on 
that parameter regardless of the values assumed for the others. 
Conversely, $1/\sqrt{\left(W^{-1}\right)_{ii}}$ determines the value
of the intersection of the ellipsoid with the corresponding $i$-parameter 
axis, and represents the 95\% C.L. bound on that parameter assuming 
all the others to be exactly known. Accordingly, the ellipsoidal surface 
constrains, at the 95\% C.L. and model-independently, the range of values  
of the CI couplings $\epsilon_{\alpha\beta}$ allowed by the foreseen 
experimental uncertainties.    

The numerical results can be represented graphically.
As an example, in Figs.~1--2 we show the planar ellipses that are
obtained 
by projecting onto the two planes 
($\epsilon_{LL},\epsilon_{LR}$) and ($\epsilon_{LR},\epsilon_{RR}$)
the 95\% C.L. allowed three-dimensional ellipsoid.

To appreciate the significant role of initial beam polarization 
we show  in Figs.~1--2  the allowed bounds, obtained from
unpolarized initial beams $P^-=P^+=0$ (dotted line), polarized electrons
and unpolarized positrons $\vert P^-\vert=0.8$, $P^+=0$ (dashed line), 
and both polarized beams with $\vert P^-\vert=0.8$ and 
$\vert P^+\vert=0.6$ (solid line).
As can be seen from those figures, the electron polarization is obviously 
essential to dramatically reduce the allowed domains of CI parameters with
respect to the case of unpolarized beams. 
The results presented in Figs.~1--2 show that the increase in the 
sensitivity to the parameters $\epsilon_{\alpha\beta}$ due to 
the additional availability of positron polarization is quite useful.
One can obtain from Figs.~1--2 the corresponding bounds on 
$\Lambda_{\alpha\beta}$. For the case when both initial beams are polarized,
we have: $\Lambda_{RR}>38$ TeV,  $\Lambda_{LL}>38$ TeV and  
$\Lambda_{LR}>49$ TeV.

\section*{Acknowledgments}
It is a pleasure to thank Dr. A. Babich, Prof. P. Osland and Prof. 
N. Paver for the fruitful and enjoyable collaboration on the topics 
related with those covered here.
\goodbreak


\begin{thebibliography}{99}

\bibitem{Eichten}
E.~Eichten, K.~Lane and M.~E.~Peskin,
Phys.\ Rev.\ Lett.\ {\bf 50} (1983) 811.

\bibitem{Schrempp}
B.~Schrempp, F.~Schrempp, N.~Wermes and D.~Zeppenfeld,
Nucl.\ Phys.\ B {\bf 296} (1988) 1.

\bibitem{Renard}
M.~Beccaria, F.~M.~Renard, S.~Spagnolo and C.~Verzegnassi, 
Phys.\ Rev.\ D {\bf 62} (2000) 053003 [hep-ph/000210].

\bibitem{Pankov1} 
A.~A.~Pankov and N.~Paver, preprint IC/2001/125, Trieste [hep-ph/0109203].

\bibitem{Cuypers}
F.~Cuypers and P.~Gambino,
Phys.\ Lett.\ B {\bf 388} (1996) 211
[hep-ph/9606391].

\bibitem{Pankov2} 
A.~A.~Babich, P.~Osland, A.~A.~Pankov and N.~Paver, 
Phys.\ Lett.\ B.{\bf 518} (2001) 128 [hep-ph/0107159]. 
Phys.Lett.B518:128-136,2001 

\end{thebibliography}
\end{document}